\documentclass[a4paper,english,prl,twocolumn]{revtex4}
\usepackage{ae} 
\usepackage[T1]{fontenc}
\usepackage[ansinew]{inputenc}
\usepackage{amsmath}
\usepackage{amssymb}
\usepackage{amsfonts}
\usepackage[dvips]{graphicx}
\usepackage{eurosym}
\usepackage{babel}
\usepackage{multirow}

\newcommand{\tbf}{\textbf}

\newcommand{\HHp}{H$_{2}^{+}$}
\newcommand{\DDp}{D$_{2}^{+}$}
\newcommand{\HHHp}{H$_{3}^{+}$}

\newcommand{\op}{\widehat}

\renewcommand{\d}{\text{d}}
\begin{document}

\title{Hydrogen molecule ion: Path integral Monte Carlo approach}
\author{I. Kylänpää $^{*}$, M. Leino and T. T. Rantala}
\affiliation{Institute of Physics, Tampere University of Technology,
P-O. Box 692, FI-33101 Tampere, Finland} 
\date{\today}

\begin{abstract}
  Path integral Monte Carlo approach is used to study the coupled
  quantum dynamics of the electron and nuclei in hydrogen molecule
  ion.  The coupling effects are demonstrated by comparing differences
  in adiabatic Born--Oppenheimer and non-adiabatic simulations, and
  inspecting projections of the full three-body dynamics onto
  adiabatic Born--Oppenheimer approximation.
  
  Coupling of electron and nuclear quantum dynamics is clearly seen.
  Nuclear pair correlation function is found to broaden by
  $0.040~a_{0}$ and average bond length is larger by $0.056~a_{0}$.
  Also, non-adiabatic correction to the binding energy is found.
  Electronic distribution is affected less, and therefore, we could
  say that the adiabatic approximation is better for the electron than
  for the nuclei.
\end{abstract}
\maketitle

\section{Introduction}
There is a number of phenomena in molecular and chemical physics which
are influenced by the quantum behavior of both nuclei and electrons,
rovibrational dynamics being a good example, see
Refs. \cite{Marx96,Ramirez03,Ramirez06} and references therein. In
case of light-mass nuclei, protons in particular, treatment of the
quantum nature of the nuclei is essential
\cite{Cheng95,Leino2,Leino3}.  This has proven to be important in the
description of hydrogen bond, for example \cite{Tuckerman97}.

Hydrogen molecule ion (\HHp{}), being the simplest molecule, has been
studied extensively \cite{Dickinson33} and it has often been used as
an example or a test case for an improved method or accuracy
\cite{Silverman86,Adamowicz86,Jones92,Macek94,Serov02}. In addition to
the free molecule, \HHp{}{} influenced by an electric or magnetic
field is a well-studied subject
\cite{Vincke85,Babb90,Tang91,Kappes96,Jones99,Moss00,Amovilli06}.
Furthermore, there is interest in descriptions that do not restrict to
Born--Oppenheimer (BO) or other adiabatic approximations
\cite{Bhatia98,Taylor99,Korobov01,Ohta03,Gross01,Gross06}.  Such
extensions can be easily realized by using quantum Monte Carlo (QMC)
methods \cite{Traynor91,Bressanini97}, for example.

Among the QMC methods the path integral formalism (PIMC) offers a
finite-temperature approach together with a transparent tool to trace
the correlations between the particles involved.  Though
computationally extremely demanding, with some approximations it is
capable of treating low-dimensional systems, such as small molecules
or clusters accurately enough.  Some examples found in literature are
H \cite{Li87}, HD$^+$ and \HHHp{} \cite{Knoll00}, H$_2$ clusters
\cite{Surh97,Gordillo99,Gordillo02,Boninsegni04,Cuervo06} with special
attention laid on $^4$He
\cite{Abraham87,Ceperley95,Pierce98,Pierce99,Kwon99}.  The
approximations in these approaches relate to the \emph{ad hoc} type
potentials describing the interactions between particles.

In this work we evaluate the density matrix of the full three-body
quantum dynamics in a stationary state and finite-temperature.
This is what we call "all-quantum" (AQ) simulation.  Secondly, the
electronic part only is evaluated as a function of internuclear
distance in the spirit of BO approximation, and thirdly, the adiabatic
nuclear dynamics is evaluated in the BO potential curve.  These allow
us to demonstrate the non-adiabatic electron--nuclei coupling by a
projection of the AQ dynamics onto the adiabatic approximations.

We need to approximate the $-1/r$ Coulomb potential of
electron--nucleus interaction at short range to make calculations
feasible.  We realize this with a carefully tested pseudopotential
(PP).  Also, the absent (ortho) or negligible (para) exchange
interaction of nuclei is not taken into account. Finally, we have to
simulate a finite temperature mixed state.  For convenience, we have
chosen $300$ K, but this essentially restricts the system to its
electronic ground state.

We begin with a brief introduction to the theory and methods in the
next section.  This includes description of the PP, and tools and
concepts for the analysis in the following section.  Then we carry on
to the results.

\section{Theory and Methods}
For a quantum many-body system in thermal equilibrium the partition
function contains all the information of the system \cite{Kleinert}.
The local thermodynamical properties, however, are included in the
density matrix from which all the properties of the quantum system may
be derived \cite{Pol87}. The non-adiabatic effects are directly taken
into account in PIMC. In addition, finite temperature and correlation
effects are exactly included.

\subsection{Path integral Monte Carlo approach}
According to the Feynman formulation of the statistical quantum
mechanics \cite{Fey72} the partition function for interacting
distinguishable particles is given by the trace of the density matrix,
\begin{align}
Z 
& = \text{Tr}~\hat{\rho}(\beta)\nonumber\\ 
& = \lim_{M\rightarrow\infty}\int \d R_{0}\d R_{1}\d R_{2} \ldots
\d R_{M-1} \prod_{i = 0}^{M-1}e^{-S(R_{i},R_{i+1};\tau)},
\end{align}
where $\hat{\rho}(\beta) = e^{-\beta\hat{H}}$, $S$ is the action,
$\beta = 1/k_{\text{B}}T$, $\tau = \beta/M$ and $R_{M}=R_{0}$. $M$ is
called the Trotter number and it characterizes the accuracy of the
discretized path. In the limit $M\to\infty$ we are ensured to get the
correct partition function $Z$, but in practice sufficient convergence at
some finite $M$ is found, depending on the steepness of the
Hamiltonian $\hat H$.

In the primitive approximation scheme of the PIMC formalism the action
is written as \cite{Cep95}
\begin{align}
S(R_{i},R_{i+1};\tau) 
= & \frac{3N}{2} \ln (4\pi\lambda\tau)
+ \frac{(R_{i}-R_{i+1})^{2}}{4\lambda\tau}\nonumber\\ 
&+ U(R_{i},R_{i+1};\tau),
\end{align}
where $U(R_{i},R_{i+1};\tau) = \frac{\tau}{2}[V(R_{i})+V(R_{i+1})]$
and $\lambda = \hbar^{2}/ 2m$. 

Sampling of the configuration space is carried out using the
Metropolis procedure \cite{Metro53} with the bisection moves
\cite{Chakravarty98}. This way the kinetic part of the action is
sampled accurately and only the interaction part is needed in the
Metropolis algorithm.  Level of the bisection sampling ranges from 3
to 6 in our simulations, respectively with the increase in the Trotter
number.  The bisection sampling turns out to be essential with large
Trotter numbers to achieve feasible convergence, for nuclei in
particular. Total energy is calculated using the virial estimator
\cite{Herman82}.

\subsection{Extrapolation of expectation values}

The Trotter scaling procedure \cite{Knoll00} for expectation values is
used to obtain estimates for energetics in the limit
$M\rightarrow\infty$. To use this procedure one needs expectation
values with several different Trotter numbers. For the Trotter number
$M$ the scaling scheme is
\begin{align}
\label{eq:extrap}
\langle\op{A}\rangle_{\infty} = \langle\op{A}\rangle_{M}
+ \sum_{i=1}^{N}\frac{c_{2i}}{M^{2i}},
\end{align}
where coefficients $c_{2i}$ are constants for a given temperature and
$N$ represents the order of extrapolation. In this paper $N=2$ has
been used for the energies of \HHp{}, and $N=3$ for hydrogen atom
energies, see Figs.~\ref{fig:Etot_H} and \ref{fig:potkayrat2}.

\subsection{Pseudopotential of the electron}
For the hydrogen molecule ion the potential energy is
\begin{align}
\label{eq:pot}
V(\tbf{r}_{1},\tbf{r}_{2},\tbf{R}) 
=  -\frac{1}{r_{1}}
-\frac{1}{r_{2}} +\frac{1}{R},
\end{align}
where $r_{i} = |\tbf{r}-\tbf{R}_{i}|$, $R =
|\tbf{R}_{1}-\tbf{R}_{2}|$, $\tbf{r}$ being the coordinates of the
electron and $R$ the internuclear distance. Eq.~\eqref{eq:pot} sets
challenges for PIMC arising from the singularity of the attractive
Coulomb interaction \cite{Thijssen,Ivanov03}, which in this work is
replaced by a PP of the form \cite{PP}
\begin{align}
\label{eq:PP}
V_{\text{PP}}(r) =
-\frac{\text{erf}(\alpha_{c}r)}{r}+(a+br^{2})e^{-\alpha r^{2}}.
\end{align}

The parametres $\alpha_{c} = 3.8638$, $\alpha = 7.8857$, $a = 1.6617$
and $b =-18.2913$ were fitted using direct numerical solution to give
the exact ground state energy of hydrogen atom and the wave function
accurately outside a cut-off radius of about $0.6~a_{0}$. Also, a
number of lowest energy orbitals of the hydrogen atom are obtained
accurately outside the same cut-off radius \cite{Ile06}.  Because the
bond length of \HHp{} is about $2~a_{0}$, it is expected that
bonding of the hydrogen molecule ion becomes properly described.

\begin{center}
\begin{figure}[tb]
\includegraphics[bb =102    45   380   270, width=0.45\textwidth]{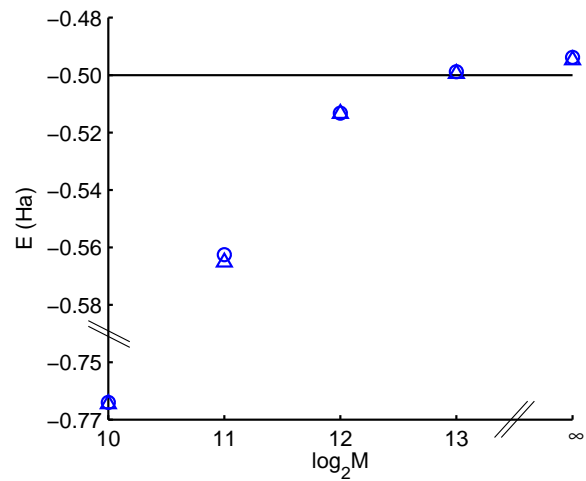}
\caption{\label{fig:Etot_H}(Color online) Hydrogen atom total energies
with different Trotter numbers: infinite nuclear mass (triangle) and
AQ (circle).  Extrapolated ground state energies are $-0.4947(1)$ Ha
and $-0.4938(3)$ Ha for infinite nuclear mass and AQ simulations,
respectively.}
\end{figure}
\end{center}

Hydrogen atom reference energies for different Trotter numbers are
shown in Fig~\ref{fig:Etot_H}, where triangles are obtained from
infinite nuclear mass and circles are from AQ simulations.
Extrapolated ground state values are $-0.4947(1)$ Ha and $-0.4938(3)$
Ha for infinite nuclear mass and AQ simulations, respectively,
statistical standard error of mean (SEM) given as uncertainty in
parenthesis.  We can note that within $2$SEM limits proportion of
these energies $0.9982$ reproduces that of Rydberg constants,
$R_{\text{H}}/R_{\infty} = 0.9995$.

\subsection{Spectroscopic constants}

Within the BO approximation of diatomic molecules the corrections to
electronic energies due to rovibrational motion of the nuclei can be
evaluated from a Dunham polynomial \cite{Alexander05}
\begin{align}
\label{eq:rovib}
E_{vJ} =& -D_{e} + \omega_{\text{e}}(v+\frac{1}{2})
- \omega_{\text{e}}x_{\text{e}}(v+\frac{1}{2})^{2}\nonumber\\
&+ B_{\text{e}}J(J+1)
- \alpha_{\text{e}}J(J+1)(v+\frac{1}{2}) + \ldots,
\end{align}
where $v$ and $J$ are vibrational and rotational quantum numbers,
respectively, and $B_{\text{e}}$, $\omega_{\text{e}}$,
$\omega_{\text{e}}x_{\text{e}}$ and $\alpha_{\text{e}}$ are the
spectroscopic constants.

The spectroscopic constants of \HHp{} and \DDp{} are obtained as
introduced in Ref. \cite{Alexander05}. In atomic
units
\begin{align}
\label{eq:Be}
&
B_{\text{e}} = \frac{1}{2 \text{I}} = \frac{1}{2\mu R^{2}},\\
\label{eq:we}
&
\omega_{\text{e}} = \Big( \frac{1}{\mu}\frac{\d^{2}E}{\d R^{2}}\Big)^{1/2},\\
\label{eq:wexe}
&
\omega_{\text{e}}x_{\text{e}} = 
\frac{1}{48\mu}\Big[5
\Big(\frac{\d^{3}E/\d R^{3}}{\d^{2}E/\d R^{2}}\Big)^{2}
-3
\frac{\d^{4}E/\d R^{4}}{\d^{2}E/\d R^{2}}\Big]\\
& \text{and}\nonumber\\
\label{eq:ae}
&
\alpha_{\text{e}} = -\frac{6B_{\text{e}}^{2}}{\omega_{\text{e}}}
\Big[\frac{R}{3}\frac{\d^{3}E/\d R^{3}}{\d^{2}E/\d R^{2}} + 1\Big].
\end{align}
Instead of determining these constants at the equilibrium distance
only, as in Ref.~\cite{Alexander05}, we evaluate expectation values
from the distribution of nuclei, e.g. for the rotational constant,
\begin{align}
\label{eq:weighting}
  B_{\text{e}} = 
  \frac{1}{2\mu} \int g(R)\frac{1}{R^{2}}\d R, 
\end{align}
where the pair correlation function $g(R)$ is normalized to unity.
The other constants, Eqs. \eqref{eq:we}--\eqref{eq:ae}, are evaluated
similarly.

\subsection{Centrifugal distortion}

Effects caused by the centrifugal distortion, arising from rotational
motion of the nuclei, on the equilibrium distance can be assessed by
inspecting the extremum values of the energy of harmonic oscillator in
rotational motion: $E_{J}(r) = \tfrac12 k(r-r_{\text{e}})^{2}+
J(J+1)/2\mu r^{2}$. We find an approximate equation
\begin{align}
\label{cfd}
\Delta R 
= \frac{4B_{\text{e}}}{\mu\omega_{\text{e}}^{2}R_{\text{e}}^{2}}J(J+1),
\end{align}
where $R_{\text{e}}$ is the equilibrium distance. Eq.~\eqref{cfd},
however, does not include the anharmonic effects shown in
Eq.~\eqref{eq:rovib}, which evidently increase the bond length.

At finite temperature the rotational energy states should be
weighted by the Boltzmann factor, which leads to
\begin{align}
\label{eq:cfd}
\Delta R
= 
\frac{4B_{\text{e}}}{\mu\omega_{\text{e}}^{2}R_{\text{e}}^{2}}
\frac{\sum_{J}J(J+1)\exp(-\beta B_{\text{e}}J(J+1))}
     {\sum_{J}\exp(-\beta B_{\text{e}}J(J+1))},
\end{align}
where $J = 0,~1,~2,\ldots$. Using the spectroscopic constants from
Ref.~\cite{Alexander05}, see Table~\ref{table:spect}, and temperature
of $300$ K we obtain $\Delta R = 0.0043~a_{0}$. This approximation will
be compared to our direct evaluation, below.

\section{Results}
We consider three different cases separately in order to demonstrate
the non-adiabatic effects.  First, the electronic part only is
evaluated as a function of internuclear distance in the spirit of BO
approximation. Secondly, the adiabatic nuclear dynamics is evaluated
in the BO potential curve. Finally, \HHp{} is treated fully
non-adiabatically with the AQ simulation. These allow us to
demonstrate the non-adiabatic electron--nuclei coupling by a
projection of the AQ dynamics onto the adiabatic approximations. In
addition, spectroscopic constants and isotope effects are looked into.

\subsection{Adiabatic electron dynamics}
\begin{figure}[t]
\includegraphics[bb = 98 45 373 272, width=0.45\textwidth]{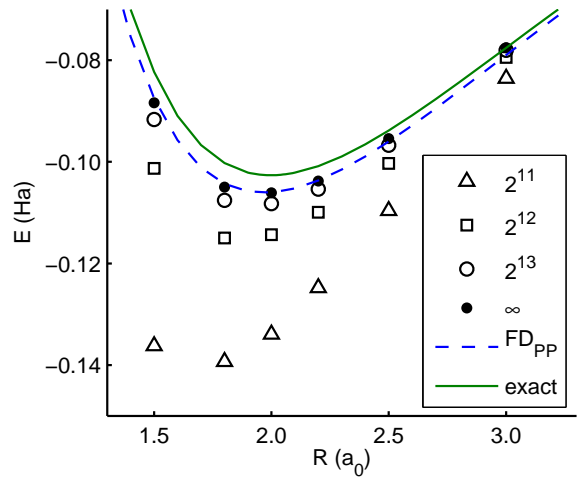}
\caption{\label{fig:potkayrat2}(Color online) \HHp{} potential curves
with different Trotter numbers: $M = 2^{11}$ (square), $M = 2^{12}$
(triangle), $M=2^{13}$ (circle), extrapolated values (dot), finite
difference calculation with the pseudopotential (dashed) and with
exact $e^{-}$--$p^{+}$ potential (solid).}
\end{figure}

Though the PP, Eq.~\eqref{eq:PP}, reproduces the hydrogen atom energy
exactly, an error of $-0.00342$ Ha from the exact value $-0.10263$ Ha
results in binding of another proton to form \HHp{}. This is
demonstrated in Fig.~\ref{fig:potkayrat2}, where potential curves of
\HHp{} from finite difference calculations with $V_{\text{PP}}$ from
Eq.~\eqref{eq:PP} and exact $V(r) = -r^{-1}$ are shown.

Our PIMC energies with increasing Trotter number $M$ and the
extrapolation to $M=\infty$ using Eq.~\eqref{eq:extrap} are shown in
the same figure. These indicate clearly that the Trotter number has to
be at least $2^{13}$ in order to find the minimum of the potential
curve at the nuclear separation $R = 2.0~a_{0}$. The extrapolated
values are in good agreement with the potential curve
FD$_{\text{PP}}$, and there is almost a perfect match at $R =
2.0~a_{0}$, where the value of the extrapolated dissociation energy is
$0.1061(2)$ Ha.

For larger nuclear separations than $3.5~a_{0}$, however, we are not
able to reproduce the potential curve with these Trotter numbers: we
get too weakly binding molecule. This is assumed to be a consequence of
the electronic wave function becoming more delocalized as the
internuclear distance increases, and thus the ''polymer ring''
representing the electron is not capable of sufficient sampling of
configuration space.  This error should diminish with increasing
$M$. 

The electron--nucleus pair correlation function is shown in
Fig.~\ref{fig:en_pk} and will be discussed below.

\begin{table}[tb]
\caption{\label{table:spect}Expectation values of spectroscopic
constants, Eqs.~\eqref{eq:Be}--\eqref{eq:weighting}.
A Morse potential \cite{Morse29} fitted to the FD$_{\text{PP}}$
potential curve is used in the evaluation of the energy
derivatives. Corresponding pair correlation functions are shown in
Fig.~\ref{fig:pkt}. First two columns are adiabatic nuclear dynamics
results and AQ results are in the third column.}
\begin{tabular}{cccccc}
\hline\hline 
 & \multicolumn{2}{c}{\underline{\hspace{0.4cm}\HHp{}\hspace{0.4cm}}}
 & \underline{\hspace{0.2cm}\DDp{}\hspace{0.2cm}}
 & \underline{\hspace{0.2cm}\HHp{} (AQ)\hspace{0.2cm}}
 & \\
 & Ha & cm$^{-1}$ & cm$^{-1}$ & cm$^{-1}$ &\\
\hline
\multirow{2}{*}{$B_{\text{e}}$} &  $0.0001366$  & $30.35$ 
               &  $15.24$      & $29.26$ & This work\\
               &  $0.0001344$  & $29.85705$ & & & Ref. \cite{Alexander05}\\
\multirow{2}{*}{$\omega_{\text{e}}$} & $0.0104816$ & $2328.96$ 
                    & $1668.25$   & $2229.77$& This work\\
 & $0.0104201$ & $2315.3$  & & ($2232$)\footnotemark[1] & 
\cite{Alexander05}, \cite{Gross06}\footnotemark[1]\\
\multirow{2}{*}{ $\omega_{\text{e}}x_{\text{e}}$} &
   $0.0003552$ & $78.92$ & $35.33$ & $90.73$ & This work\\
 & $0.0003029$ & $67.3$  & & & Ref. \cite{Alexander05}\\
\multirow{2}{*}{$\alpha_{\text{e}}$} &
   $6.445\times 10^{-6}$ & $1.432$ & $0.45$ & $1.636$ & This work\\
 & $7.201\times 10^{-6}$ & $1.600$ & & & Ref. \cite{Alexander05}\\
\hline\hline  
\end{tabular}
\end{table}

\subsection{Adiabatic nuclear dynamics}
\begin{figure}[tb]
\includegraphics[bb=84    52   293   286,width=0.35\textwidth]{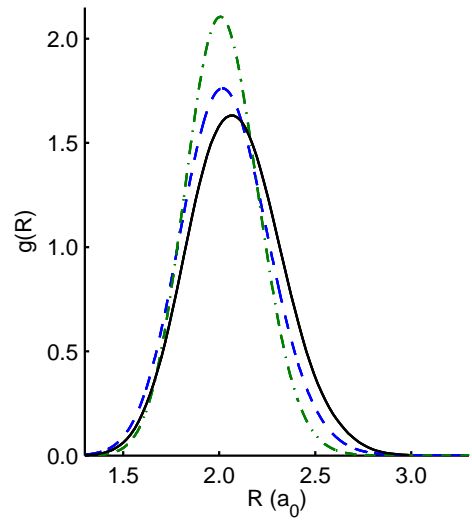}
\caption{\label{fig:pkt}(Color online) Nuclear pair correlation
functions: \HHp{} AQ (solid), \HHp{} QN (dashed) and \DDp{} QN
(dash-dotted).  The difference in the average nuclear separation
between QN and AQ \HHp{} is $0.056(3)~a_{0}$.}
\end{figure}

For the quantum dynamics of the nuclei only (QN) we consider both
\HHp{} and \DDp{} to see the isotope effect, too. The FD$_{\text{PP}}$
potential curve in Fig.~\ref{fig:potkayrat2} is used, for which
convergence with respect to Trotter number is found at $M\geq 2^{6}$
for both isotopes.  Resulting pair correlation functions are shown in
Fig.~\ref{fig:pkt}.

Average nuclear separation of $2.019(1)~a_{0}$ for \HHp{} and
$2.007(2)~a_{0}$ for the isotope \DDp{} is found with $M\geq 2^{6}$.
The full width at half maximum (FWHM) of the pair correlation
functions are $0.539(1)~a_{0}$ and $0.454(1)~a_{0}$ for these
isotopes, respectively.

Difference in the bond length of \HHp{} between the adiabatic electron
and adiabatic nuclei simulations, i.e total distortion, is
$0.019~a_{0}$. Centrifugal contribution to this, the difference
between one and three dimensional simulations of the nuclei, is
$0.009(1)~a_{0}$, which unexpectedly is about twice as much as the
value $0.0043~a_{0}$ evaluated from the approximate
Eq.~\eqref{eq:cfd}. The anharmonic contribution, i.e. difference
between total and centrifugal distortions, is $0.010(1)~a_{0}$. In
Ref.~\cite{Lounila91} it was shown that anharmonic effects in H$_2$
molecule contribute about the same amount to total distortion as
centrifugal force, which turns out to be the case here, too.

Difference between the total energies of the previous simulations (3D
vs. 1D) is $0.0009383(2)$ Ha, which is close to $k_{\text{B}}T\approx
0.00095$ Ha as expected due to the presence of the two rotational
degrees of freedom in 3D. Difference between the dissociation energies
of adiabatic electron and nuclear simulations, i.e. the zero-point
vibrational energy, is $0.0064(2)$ Ha.

A Morse potential \cite{Morse29} fitted to the FD$_{\text{PP}}$
potential curve is used in the evaluation of the spectroscopic
constants, see Table~\ref{table:spect}. This is justified because the
nuclear simulations and analytical Morse wave function \cite{terHaar}
calculations coincide.  The spectroscopic constants of \HHp{} are
close to those given in Ref. \cite{Alexander05}, which have been
determined at the equilibrium distance of the nuclei, only.  Same
procedure is used for the spectroscopic constants of the other
isotope. In Table~\ref{table:spect} the same constants evaluated using
the AQ instead of BO nuclear pair correlation function are
also shown.

\subsection{Non-adiabatic ''all-quantum'' dynamics}
For \HHp{} the total energy of AQ simulation with the Trotter number
$M=2^{13}$ is $-0.60159(3)$ Ha. The extrapolation procedure yields
total energy $-0.59872(3)$ Ha, which is only $0.0016$ Ha more binding
than the value $-0.5971$ Ha from variational Monte Carlo (VMC)
simulation \cite{Bressanini97}. The zero-point energy obtained from
simulations is $D_{e}-D_{0}^{0}=0.0074$ Ha, see Table
\ref{table:DeAQ}. It should be pointed out that the error due to the
pseudopotential in the AQ total energy is only about half of that
found for the BO total energies.

Difference in dissociation energies of AQ and the 3D QN \HHp{}
simulations is $0.00097$ Ha, which is about $k_{\text{B}}T$ revealing
additional electronic energy degrees of freedom in the first. AQ
simulation for \HHp{} gives for the average nuclear separation $R =
2.075(2)~a_{0}$, which is $0.056~a_{0}$ larger than that in the QN
simulation. The AQ FWHM of the nuclear pair correlation function is
$0.5785(2)~a_{0}$, which shows a spreading of $0.040~a_{0}$ compared
to the QN results, see Fig.~\ref{fig:pkt}.

\begin{figure}[tb]
\includegraphics[bb=84 52 300 291, width=0.35\textwidth]{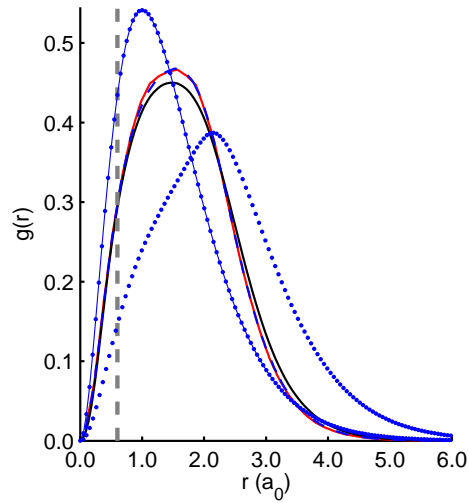}
\caption{\label{fig:en_pk}(Color online) \HHp{} electron--nucleus pair
correlation functions: AQ (solid, second lowest curve), AQ projection
to $R\approx 2.0~a_{0}$ (solid) and BO at $R=2.0~a_{0}$ (dashed). The
latter two almost coincide. Dashed vertical line indicates the size of
the pseudopotential core, $r = 0.6~a_{0}$. For comparison
corresponding pair correlation functions for hydrogen atom (dotted
line) and \HHp{} (dotted) obtained by using the analytical ground
state wave function of hydrogen atom are also shown.}
\end{figure}

In Fig.~\ref{fig:en_pk} BO and AQ electron--nucleus pair correlation
functions are compared. AQ projection onto the BO bond length, $R =
2.0~a_{0}$, and BO results coincide, which indicates that the
adiabatic BO approach for the electron dynamics is sufficient. Thus,
it seems that the electron--nuclei coupling effects are more clearly
seen in the dynamics of the nuclei, see Fig.~\ref{fig:pkt}.  As one
might expect, there is a noticeable difference between the
AQ and the BO electron--nucleus pair correlation
functions due to varying bond length, see Fig.~\ref{fig:en_pk}.

\begin{table}[tb]
\begin{center}
\caption{\label{table:DeAQ}H$_2^{+}$ energetics (atomic units). First
three are BO and the next three are non-adiabatic values.}
\begin{tabular}{p{1.8cm}cccc}
\hline\hline
 Method & $E_{\text{tot}}$ & $D_{e}$ & $D_{0}^{0}$ & $R$\\ 
\hline
HF\footnotemark[1] & $-0.6026$ & $0.1026$ & & $2.000$\\ 
VMC\footnotemark[2] & $-0.6026$ & $0.1026$ & & $2.000$\\
PIMC\footnotemark[5] & $-0.6061(2)$ & $0.1061(2)$ & $0.0997(1)$ & $2.0$\\ 
\hline
VMC\footnotemark[3] & $-0.5971$ & & $0.0971$ & $2.064$\\
MCDFT\footnotemark[4] & $-0.581$ & & $0.081$ & $2.08$\\
PIMC\footnotemark[5] & $-0.59872(3)$ & & $0.09872(3)$ & $2.075(2)$\\
\hline\hline
\end{tabular}
\footnotetext[1]{\cite{Pyykko}: Hartree--Fock}
\footnotetext[2]{\cite{Alexander05}: VMC, Born--Oppenheimer} 
\footnotetext[3]{\cite{Bressanini97}: VMC, non-adiabatic} 
\footnotetext[4]{\cite{Gross06}: MCDFT, non-adiabatic ({\it SAO})}
\footnotetext[5]{This work}
\end{center}
\end{table}

The AQ average nuclear separation is close to the value $2.064~a_{0}$
obtained by a non-adiabatic VMC simulation \cite{Bressanini97}. The
AQ pair correlation function of the nuclei, see Fig.~\ref{fig:pkt},
coincides with the {\it SAO} (Scaled Atomic Orbital) one in
Ref.~\cite{Gross06} computed within the Multicomponent Density
Functional Theory (MCDFT) scheme, not shown here.

All the spectroscopic constants in Table~\ref{table:spect} are defined
using the derivatives from a fitted Morse potential, i.e. BO potential
energy surface. Thus, the ''AQ spectroscopic constants'' should be
interpreted mainly as the direction of change in the values, except
for $B_{\text{e}}$.  The expectation values of the spectroscopic
constants are obtained by weighting the equations by the nuclear pair
correlation function from the corresponding simulation.

\begin{figure}[tb]
\includegraphics[bb=98 45 372 270, width=0.45\textwidth]{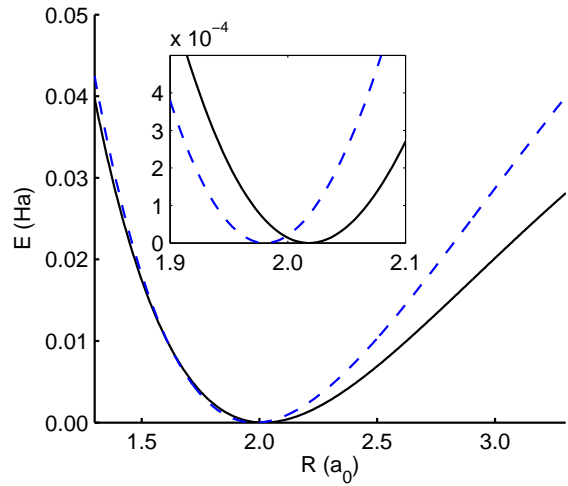}
\caption{\label{fig:potBOAQ}(Color online) \HHp{} potential curves:
Morse potential fitted to FD$_{\text{PP}}$ (dashed) and the effective
Morse potential obtained from the projection of the AQ simulation
(solid), see the text for details. Corresponding nuclear pair
correlation functions are shown in Fig.~\ref{fig:pkt}. The shift in
the bond length is $0.036~a_{0}$.}
\end{figure}

A projection of the AQ simulation to a potential curve of the nuclei
is constructed with the help of the known solutions to the Morse
potential. Distribution from the Morse wave function is fitted to the
pair correlation function of the AQ simulation. The three-body system
is then presented by an effective two-body potential. The projected
potential curve shows clear differences in the dynamics of the nuclei
between BO and AQ simulations, see Fig.~\ref{fig:potBOAQ}. The minima
of the potentials are set to zero: the difference in the
dissociation energies between BO and the AQ projection is about
$0.036$ Ha and the shift in the equilibrium distance is
$0.036~a_{0}$. The spectroscopic constants with the projected
potential curve are $B_{\text{e}} = 29.26~$cm$^{-1}$,
$\omega_{\text{e}} = 2047.94~$cm$^{-1}$,
$\omega_{\text{e}}x_{\text{e}} = 78.12~$cm$^{-1}$ and
$\alpha_{\text{e}} = 2.110~$cm$^{-1}$.  All this indicates that an
effective Morse potential is not capable of describing non-adiabatic
effects correctly.

\begin{figure}[tb]
\includegraphics[bb = 98 45 372 265, width=0.45\textwidth]{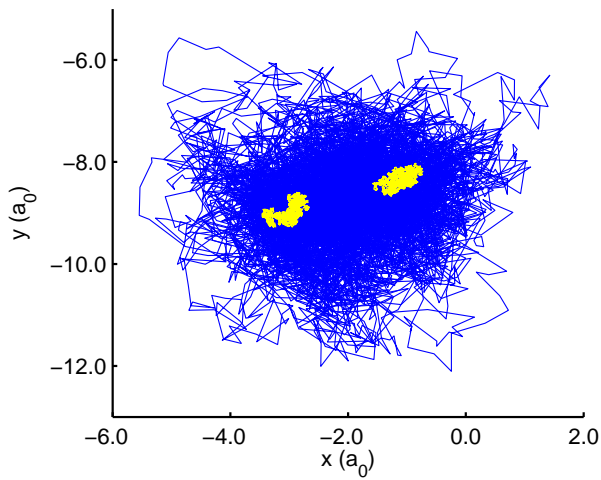}
\caption{\label{fig:xy_taso}(Color online) xy-plane (z-projection)
snapshot from AQ simulation with Trotter number $2^{13}$ for all
particles. ''Polymer ring'' describing the electron is in the
background and those of the nuclei are placed on top.}
\end{figure}

Finally, it may be of interest to see a visualization of the ''polymer
rings'' representing the quantum particles in the PIMC simulation.
So, Fig.~\ref{fig:xy_taso} presents the xy-plane (z-projection)
snapshot from AQ simulation with Trotter number $2^{13}$ for all three
particles. ''Polymer ring'' describing the electron is in the
background and those of the nuclei are placed on top.

\section{Conclusions}
The three-body quantum system, hydrogen molecule ion (\HHp{}), is
revisited, once again. Path integral Monte Carlo (PIMC) method is used
for evaluation of the stationary state quantum dynamics. PIMC offers a
finite-temperature approach together with a transparent tool to
describe the correlations between the particles involved. We aim at
tracing the electron--nuclei coupling effects in the three-body
all-quantum (AQ), i.e. non-adiabatic, molecule.  This is carried out
by comparing the differences in adiabatic Born--Oppenheimer (BO) and
AQ simulations, and inspecting the projections from the AQ simulation
onto the BO description of the electron-only and nuclear-only
subsystems.

The approach turns out to be computationally demanding, but with the
chosen pseudopotential for the attractive Coulomb potential and
extrapolation to infinite Trotter number the task becomes feasible. By
choosing low enough temperature, $300$ K, we are able to compare our
data to those from zero--Kelvin quantum methods available in
literature. Among others we have evaluated spectroscopic constants
and molecular deformation, also considering the isotope effects.

With our fully basis set free, trial wave function free and model free
approach we are not able to compete in accuracy with the zero--Kelvin
benchmark values. However, due to the mixed state density matrix
formalism of PIMC we are able to present the most transparent
description of the particle--particle correlations.

Total energies from our simulations are more binding in nature
compared to the benchmark values, see Table~\ref{table:DeAQ}. This is
an expected effect of the pseudopotential in use, see
Fig.~\ref{fig:potkayrat2} and FD$_{\text{PP}}$ therein. Quantum
dynamics of the system is well described and distinct features of
coupling are observed for the nuclei: shift of $0.056~a_{0}$ in the
equilibrium bond length, increase of $0.040~a_{0}$ in the width of the
pair correlation function of the nuclei and non-adiabatic correction
of about $0.00097$ Ha to dissociation energy. Electronic distribution,
however, is less influenced by the coupling, see Fig.~\ref{fig:en_pk},
and therefore, we could say that the adiabatic approximation is better
for the electron than for the nuclei.

Projection of the non-adiabatic three-body system with the help of
Morse wave functions onto two-body nuclei-only subsystem indicates
that Morse potential is not capable of describing non-adiabatic
effects correctly, see Fig.~\ref{fig:potBOAQ}.

\section{Acknowledgements}
For financial support we thank Graduate School of Tampere University
of Technology and the Academy of Finland, and for computational
resources the facilities of Finnish IT Center for Science (CSC) and
Material Sciences National Grid Infrastructure (M-grid, akaatti).

\bibliography{viitteet}

\begin{thebibliography}{56}
\expandafter\ifx\csname natexlab\endcsname\relax\def\natexlab#1{#1}\fi
\expandafter\ifx\csname bibnamefont\endcsname\relax
  \def\bibnamefont#1{#1}\fi
\expandafter\ifx\csname bibfnamefont\endcsname\relax
  \def\bibfnamefont#1{#1}\fi
\expandafter\ifx\csname citenamefont\endcsname\relax
  \def\citenamefont#1{#1}\fi
\expandafter\ifx\csname url\endcsname\relax
  \def\url#1{\texttt{#1}}\fi
\expandafter\ifx\csname urlprefix\endcsname\relax\def\urlprefix{URL }\fi
\providecommand{\bibinfo}[2]{#2}
\providecommand{\eprint}[2][]{\url{#2}}

\bibitem[{\citenamefont{Marx and Parrinello}(1996)}]{Marx96}
\bibinfo{author}{\bibfnamefont{D.}~\bibnamefont{Marx}} \bibnamefont{and}
  \bibinfo{author}{\bibfnamefont{M.}~\bibnamefont{Parrinello}},
  \bibinfo{journal}{J. Chem. Phys.} \textbf{\bibinfo{volume}{104}},
  \bibinfo{pages}{4077} (\bibinfo{year}{1996}).

\bibitem[{\citenamefont{Lopez-Ciudad et~al.}(2003)\citenamefont{Lopez-Ciudad,
  Ram\`irez, Schulte, and B\"ohm}}]{Ramirez03}
\bibinfo{author}{\bibfnamefont{T.}~\bibnamefont{Lopez-Ciudad}},
  \bibinfo{author}{\bibfnamefont{R.}~\bibnamefont{Ram\`irez}},
  \bibinfo{author}{\bibfnamefont{J.}~\bibnamefont{Schulte}}, \bibnamefont{and}
  \bibinfo{author}{\bibfnamefont{M.~C.} \bibnamefont{B\"ohm}},
  \bibinfo{journal}{J.~Chem.~Phys.} \textbf{\bibinfo{volume}{119}},
  \bibinfo{pages}{4328} (\bibinfo{year}{2003}).

\bibitem[{\citenamefont{Ram\`irez et~al.}(2006)\citenamefont{Ram\`irez,
  Herrero, and Hern\`andez}}]{Ramirez06}
\bibinfo{author}{\bibfnamefont{R.}~\bibnamefont{Ram\`irez}},
  \bibinfo{author}{\bibfnamefont{C.~P.} \bibnamefont{Herrero}},
  \bibnamefont{and} \bibinfo{author}{\bibfnamefont{E.~R.}
  \bibnamefont{Hern\`andez}}, \bibinfo{journal}{Phys.~Rev.~B}
  \textbf{\bibinfo{volume}{73}}, \bibinfo{pages}{1} (\bibinfo{year}{2006}).

\bibitem[{\citenamefont{Cheng et~al.}(1995)\citenamefont{Cheng, Barnett, and
  Landman}}]{Cheng95}
\bibinfo{author}{\bibfnamefont{H.-P.} \bibnamefont{Cheng}},
  \bibinfo{author}{\bibfnamefont{R.~N.} \bibnamefont{Barnett}},
  \bibnamefont{and} \bibinfo{author}{\bibfnamefont{U.}~\bibnamefont{Landman}},
  \bibinfo{journal}{Chem.~Phys.~Lett.} \textbf{\bibinfo{volume}{237}},
  \bibinfo{pages}{161} (\bibinfo{year}{1995}).

\bibitem[{\citenamefont{Leino et~al.}(2006)\citenamefont{Leino, Nieminen, and
  Rantala}}]{Leino2}
\bibinfo{author}{\bibfnamefont{M.}~\bibnamefont{Leino}},
  \bibinfo{author}{\bibfnamefont{J.}~\bibnamefont{Nieminen}}, \bibnamefont{and}
  \bibinfo{author}{\bibfnamefont{T.~T.} \bibnamefont{Rantala}},
  \bibinfo{journal}{Surface Science} \textbf{\bibinfo{volume}{600}},
  \bibinfo{pages}{1860} (\bibinfo{year}{2006}).

\bibitem[{\citenamefont{Leino et~al.}(2007)\citenamefont{Leino, Kyl\"anp\"a\"a,
  and Rantala}}]{Leino3}
\bibinfo{author}{\bibfnamefont{M.}~\bibnamefont{Leino}},
  \bibinfo{author}{\bibfnamefont{I.}~\bibnamefont{Kyl\"anp\"a\"a}},
  \bibnamefont{and} \bibinfo{author}{\bibfnamefont{T.~T.}
  \bibnamefont{Rantala}}, \bibinfo{journal}{Surface Science}
  \textbf{\bibinfo{volume}{601}}, \bibinfo{pages}{1246} (\bibinfo{year}{2007}).

\bibitem[{\citenamefont{Tuckerman et~al.}(1997)\citenamefont{Tuckerman, Marx,
  Klein, and Parrinello}}]{Tuckerman97}
\bibinfo{author}{\bibfnamefont{M.~E.} \bibnamefont{Tuckerman}},
  \bibinfo{author}{\bibfnamefont{D.}~\bibnamefont{Marx}},
  \bibinfo{author}{\bibfnamefont{M.~L.} \bibnamefont{Klein}}, \bibnamefont{and}
  \bibinfo{author}{\bibfnamefont{M.}~\bibnamefont{Parrinello}},
  \bibinfo{journal}{Science} \textbf{\bibinfo{volume}{275}},
  \bibinfo{pages}{817} (\bibinfo{year}{1997}).

\bibitem[{\citenamefont{Dickinson}(1933)}]{Dickinson33}
\bibinfo{author}{\bibfnamefont{B.~N.} \bibnamefont{Dickinson}},
  \bibinfo{journal}{J.~Chem.~Phys.} \textbf{\bibinfo{volume}{1}},
  \bibinfo{pages}{317} (\bibinfo{year}{1933}).

\bibitem[{\citenamefont{Silverman et~al.}(1986)\citenamefont{Silverman, Bishop,
  and Pipin}}]{Silverman86}
\bibinfo{author}{\bibfnamefont{J.~N.} \bibnamefont{Silverman}},
  \bibinfo{author}{\bibfnamefont{D.~M.} \bibnamefont{Bishop}},
  \bibnamefont{and} \bibinfo{author}{\bibfnamefont{J.}~\bibnamefont{Pipin}},
  \bibinfo{journal}{Phys.~Rev.~Lett.} \textbf{\bibinfo{volume}{56}},
  \bibinfo{pages}{1358} (\bibinfo{year}{1986}).

\bibitem[{\citenamefont{Adamowicz and Bartlett}(1986)}]{Adamowicz86}
\bibinfo{author}{\bibfnamefont{L.}~\bibnamefont{Adamowicz}} \bibnamefont{and}
  \bibinfo{author}{\bibfnamefont{R.~J.} \bibnamefont{Bartlett}},
  \bibinfo{journal}{J.~Chem.~Phys.} \textbf{\bibinfo{volume}{84}},
  \bibinfo{pages}{4988} (\bibinfo{year}{1986}).

\bibitem[{\citenamefont{Jones and Etemadi}(1992)}]{Jones92}
\bibinfo{author}{\bibfnamefont{H.~W.} \bibnamefont{Jones}} \bibnamefont{and}
  \bibinfo{author}{\bibfnamefont{B.}~\bibnamefont{Etemadi}},
  \bibinfo{journal}{Phys.~Rev.~A} \textbf{\bibinfo{volume}{47}},
  \bibinfo{pages}{3430} (\bibinfo{year}{1992}).

\bibitem[{\citenamefont{Macek and Ovchinnikov}(1994)}]{Macek94}
\bibinfo{author}{\bibfnamefont{J.~H.} \bibnamefont{Macek}} \bibnamefont{and}
  \bibinfo{author}{\bibfnamefont{S.~Y.} \bibnamefont{Ovchinnikov}},
  \bibinfo{journal}{Phys.~Rev.~A} \textbf{\bibinfo{volume}{49}},
  \bibinfo{pages}{R4273} (\bibinfo{year}{1994}).

\bibitem[{\citenamefont{Serov et~al.}(2002)\citenamefont{Serov, Joulakian,
  Pavlov, Puzynin, and Vinitsky}}]{Serov02}
\bibinfo{author}{\bibfnamefont{V.~V.} \bibnamefont{Serov}},
  \bibinfo{author}{\bibfnamefont{B.~B.} \bibnamefont{Joulakian}},
  \bibinfo{author}{\bibfnamefont{D.~V.} \bibnamefont{Pavlov}},
  \bibinfo{author}{\bibfnamefont{I.~V.} \bibnamefont{Puzynin}},
  \bibnamefont{and} \bibinfo{author}{\bibfnamefont{S.~I.}
  \bibnamefont{Vinitsky}}, \bibinfo{journal}{Phys.~Rev.~A}
  \textbf{\bibinfo{volume}{65}}, \bibinfo{pages}{1} (\bibinfo{year}{2002}).

\bibitem[{\citenamefont{Vincke and Baye}(1985)}]{Vincke85}
\bibinfo{author}{\bibfnamefont{M.}~\bibnamefont{Vincke}} \bibnamefont{and}
  \bibinfo{author}{\bibfnamefont{D.}~\bibnamefont{Baye}},
  \bibinfo{journal}{J.~Phys.~B.: At.~Mol.~Opt.~Phys.}
  \textbf{\bibinfo{volume}{18}}, \bibinfo{pages}{167} (\bibinfo{year}{1985}).

\bibitem[{\citenamefont{Babb and Dalgarno}(1990)}]{Babb90}
\bibinfo{author}{\bibfnamefont{J.~F.} \bibnamefont{Babb}} \bibnamefont{and}
  \bibinfo{author}{\bibfnamefont{A.}~\bibnamefont{Dalgarno}},
  \bibinfo{journal}{Phys.~Rev.~Lett.} \textbf{\bibinfo{volume}{66}},
  \bibinfo{pages}{880} (\bibinfo{year}{1990}).

\bibitem[{\citenamefont{Tang et~al.}(1991)\citenamefont{Tang, Toennies, and
  Yiu}}]{Tang91}
\bibinfo{author}{\bibfnamefont{K.~T.} \bibnamefont{Tang}},
  \bibinfo{author}{\bibfnamefont{J.~P.} \bibnamefont{Toennies}},
  \bibnamefont{and} \bibinfo{author}{\bibfnamefont{C.~L.} \bibnamefont{Yiu}},
  \bibinfo{journal}{J.~Chem.~Phys.} \textbf{\bibinfo{volume}{94}},
  \bibinfo{pages}{7266} (\bibinfo{year}{1991}).

\bibitem[{\citenamefont{Kappes and Schmelcher}(1996)}]{Kappes96}
\bibinfo{author}{\bibfnamefont{U.}~\bibnamefont{Kappes}} \bibnamefont{and}
  \bibinfo{author}{\bibfnamefont{P.}~\bibnamefont{Schmelcher}},
  \bibinfo{journal}{Phys.~Rev.~A} \textbf{\bibinfo{volume}{53}},
  \bibinfo{pages}{3869} (\bibinfo{year}{1996}).

\bibitem[{\citenamefont{Bouferguene et~al.}(1999)\citenamefont{Bouferguene,
  Weatherford, and Jones}}]{Jones99}
\bibinfo{author}{\bibfnamefont{A.}~\bibnamefont{Bouferguene}},
  \bibinfo{author}{\bibfnamefont{C.~A.} \bibnamefont{Weatherford}},
  \bibnamefont{and} \bibinfo{author}{\bibfnamefont{H.~W.} \bibnamefont{Jones}},
  \bibinfo{journal}{Phys.~Rev.~E} \textbf{\bibinfo{volume}{59}},
  \bibinfo{pages}{2412} (\bibinfo{year}{1999}).

\bibitem[{\citenamefont{Moss}(2000)}]{Moss00}
\bibinfo{author}{\bibfnamefont{R.~E.} \bibnamefont{Moss}},
  \bibinfo{journal}{Phys.~Rev.~A} \textbf{\bibinfo{volume}{61}},
  \bibinfo{pages}{1} (\bibinfo{year}{2000}).

\bibitem[{\citenamefont{Amovilli and March}(2006)}]{Amovilli06}
\bibinfo{author}{\bibfnamefont{C.}~\bibnamefont{Amovilli}} \bibnamefont{and}
  \bibinfo{author}{\bibfnamefont{N.~H.} \bibnamefont{March}},
  \bibinfo{journal}{Int.~J.~Quantum~Chem.} \textbf{\bibinfo{volume}{106}},
  \bibinfo{pages}{533} (\bibinfo{year}{2006}).

\bibitem[{\citenamefont{Bhatia and Drachman}(1998)}]{Bhatia98}
\bibinfo{author}{\bibfnamefont{A.~K.} \bibnamefont{Bhatia}} \bibnamefont{and}
  \bibinfo{author}{\bibfnamefont{R.~J.} \bibnamefont{Drachman}},
  \bibinfo{journal}{Phys.~Rev.~A} \textbf{\bibinfo{volume}{59}},
  \bibinfo{pages}{205} (\bibinfo{year}{1998}).

\bibitem[{\citenamefont{Taylor et~al.}(1999)\citenamefont{Taylor, Dalgarno, and
  Babb}}]{Taylor99}
\bibinfo{author}{\bibfnamefont{J.~M.} \bibnamefont{Taylor}},
  \bibinfo{author}{\bibfnamefont{A.}~\bibnamefont{Dalgarno}}, \bibnamefont{and}
  \bibinfo{author}{\bibfnamefont{J.~F.} \bibnamefont{Babb}},
  \bibinfo{journal}{Phys.~Rev.~A} \textbf{\bibinfo{volume}{60}},
  \bibinfo{pages}{R2630} (\bibinfo{year}{1999}).

\bibitem[{\citenamefont{Korobov}(2001)}]{Korobov01}
\bibinfo{author}{\bibfnamefont{V.~I.} \bibnamefont{Korobov}},
  \bibinfo{journal}{Phys.~Rev.~A} \textbf{\bibinfo{volume}{63}},
  \bibinfo{pages}{1} (\bibinfo{year}{2001}).

\bibitem[{\citenamefont{Ohta et~al.}(2003)\citenamefont{Ohta, Maki, Nagao,
  Kono, and Fujimura}}]{Ohta03}
\bibinfo{author}{\bibfnamefont{Y.}~\bibnamefont{Ohta}},
  \bibinfo{author}{\bibfnamefont{J.}~\bibnamefont{Maki}},
  \bibinfo{author}{\bibfnamefont{H.}~\bibnamefont{Nagao}},
  \bibinfo{author}{\bibfnamefont{H.}~\bibnamefont{Kono}}, \bibnamefont{and}
  \bibinfo{author}{\bibfnamefont{Y.}~\bibnamefont{Fujimura}},
  \bibinfo{journal}{Int.~J.~Quantum~Chem.} \textbf{\bibinfo{volume}{91}},
  \bibinfo{pages}{105} (\bibinfo{year}{2003}).

\bibitem[{\citenamefont{Kreibich et~al.}(2001)\citenamefont{Kreibich, {van
  Leeuwen}, and Gross}}]{Gross01}
\bibinfo{author}{\bibfnamefont{T.}~\bibnamefont{Kreibich}},
  \bibinfo{author}{\bibfnamefont{R.}~\bibnamefont{{van Leeuwen}}},
  \bibnamefont{and} \bibinfo{author}{\bibfnamefont{E.~K.~U.}
  \bibnamefont{Gross}}, \bibinfo{journal}{Phys.~Rev.~Lett.}
  \textbf{\bibinfo{volume}{86}}, \bibinfo{pages}{2984} (\bibinfo{year}{2001}).

\bibitem[{\citenamefont{Kreibich et~al.}(2006)\citenamefont{Kreibich, {van
  Leeuwen}, and Gross}}]{Gross06}
\bibinfo{author}{\bibfnamefont{T.}~\bibnamefont{Kreibich}},
  \bibinfo{author}{\bibfnamefont{R.}~\bibnamefont{{van Leeuwen}}},
  \bibnamefont{and} \bibinfo{author}{\bibfnamefont{E.~K.~U.}
  \bibnamefont{Gross}}, \emph{\bibinfo{title}{Multicomponent density-functional
  theory for electrons and nuclei}} (\bibinfo{year}{2006}),
  \urlprefix\url{http://www.citebase.org/abstract?id=oai:arXiv.org:cond-mat/06%
09697}.

\bibitem[{\citenamefont{Traynor et~al.}(1991)\citenamefont{Traynor, Anderson,
  and Boghosian}}]{Traynor91}
\bibinfo{author}{\bibfnamefont{C.~A.} \bibnamefont{Traynor}},
  \bibinfo{author}{\bibfnamefont{J.~B.} \bibnamefont{Anderson}},
  \bibnamefont{and} \bibinfo{author}{\bibfnamefont{B.~M.}
  \bibnamefont{Boghosian}}, \bibinfo{journal}{J. Chem. Phys.}
  \textbf{\bibinfo{volume}{94}}, \bibinfo{pages}{3657} (\bibinfo{year}{1991}).

\bibitem[{\citenamefont{Bressanini et~al.}(1997)\citenamefont{Bressanini,
  Mella, and Morosi}}]{Bressanini97}
\bibinfo{author}{\bibfnamefont{D.}~\bibnamefont{Bressanini}},
  \bibinfo{author}{\bibfnamefont{M.}~\bibnamefont{Mella}}, \bibnamefont{and}
  \bibinfo{author}{\bibfnamefont{G.}~\bibnamefont{Morosi}},
  \bibinfo{journal}{Chem.~Phys.~Lett.} \textbf{\bibinfo{volume}{272}},
  \bibinfo{pages}{370} (\bibinfo{year}{1997}).

\bibitem[{\citenamefont{Li and Broughton}(1987)}]{Li87}
\bibinfo{author}{\bibfnamefont{X.-P.} \bibnamefont{Li}} \bibnamefont{and}
  \bibinfo{author}{\bibfnamefont{J.~Q.} \bibnamefont{Broughton}},
  \bibinfo{journal}{J. Chem. Phys} \textbf{\bibinfo{volume}{86}},
  \bibinfo{pages}{5094} (\bibinfo{year}{1987}).

\bibitem[{\citenamefont{Knoll and Marx}(2000)}]{Knoll00}
\bibinfo{author}{\bibfnamefont{L.}~\bibnamefont{Knoll}} \bibnamefont{and}
  \bibinfo{author}{\bibfnamefont{D.}~\bibnamefont{Marx}},
  \bibinfo{journal}{Europ. Phys J. D} \textbf{\bibinfo{volume}{10}},
  \bibinfo{pages}{353} (\bibinfo{year}{2000}).

\bibitem[{\citenamefont{Surh et~al.}(1997)\citenamefont{Surh, Runge, III,
  Pollock, and Mailhiot}}]{Surh97}
\bibinfo{author}{\bibfnamefont{M.~P.} \bibnamefont{Surh}},
  \bibinfo{author}{\bibfnamefont{K.~J.} \bibnamefont{Runge}},
  \bibinfo{author}{\bibfnamefont{T.~W.~B.} \bibnamefont{III}},
  \bibinfo{author}{\bibfnamefont{E.~L.} \bibnamefont{Pollock}},
  \bibnamefont{and} \bibinfo{author}{\bibfnamefont{C.}~\bibnamefont{Mailhiot}},
  \bibinfo{journal}{Phys. Rev. B} \textbf{\bibinfo{volume}{55}},
  \bibinfo{pages}{11330(12)} (\bibinfo{year}{1997}).

\bibitem[{\citenamefont{Gordillo}(1999)}]{Gordillo99}
\bibinfo{author}{\bibfnamefont{M.~C.} \bibnamefont{Gordillo}},
  \bibinfo{journal}{Phys. Rev. B} \textbf{\bibinfo{volume}{60}},
  \bibinfo{pages}{6790} (\bibinfo{year}{1999}).

\bibitem[{\citenamefont{Gordillo and Ceperley}(2002)}]{Gordillo02}
\bibinfo{author}{\bibfnamefont{M.~C.} \bibnamefont{Gordillo}} \bibnamefont{and}
  \bibinfo{author}{\bibfnamefont{D.~M.} \bibnamefont{Ceperley}},
  \bibinfo{journal}{Phys. Rev. B} \textbf{\bibinfo{volume}{65}},
  \bibinfo{pages}{174527} (\bibinfo{year}{2002}).

\bibitem[{\citenamefont{Boninsegni}(2004)}]{Boninsegni04}
\bibinfo{author}{\bibfnamefont{M.}~\bibnamefont{Boninsegni}},
  \bibinfo{journal}{Phys. Rev. B} \textbf{\bibinfo{volume}{70}},
  \bibinfo{pages}{125405} (\bibinfo{year}{2004}).

\bibitem[{\citenamefont{Cuervo and Roy}(2006)}]{Cuervo06}
\bibinfo{author}{\bibfnamefont{J.~E.} \bibnamefont{Cuervo}} \bibnamefont{and}
  \bibinfo{author}{\bibfnamefont{P.-N.} \bibnamefont{Roy}},
  \bibinfo{journal}{J. Chem. Phys.} \textbf{\bibinfo{volume}{125}},
  \bibinfo{pages}{124314} (\bibinfo{year}{2006}).

\bibitem[{\citenamefont{Abraham and Broughton}(1987)}]{Abraham87}
\bibinfo{author}{\bibfnamefont{F.~F.} \bibnamefont{Abraham}} \bibnamefont{and}
  \bibinfo{author}{\bibfnamefont{J.~Q.} \bibnamefont{Broughton}},
  \bibinfo{journal}{Phys. Rev. Lett.} \textbf{\bibinfo{volume}{59}},
  \bibinfo{pages}{64} (\bibinfo{year}{1987}).

\bibitem[{\citenamefont{Ceperley}(1995{\natexlab{a}})}]{Ceperley95}
\bibinfo{author}{\bibfnamefont{D.~M.} \bibnamefont{Ceperley}},
  \bibinfo{journal}{Rev. Mod. Phys} \textbf{\bibinfo{volume}{67}},
  \bibinfo{pages}{279} (\bibinfo{year}{1995}{\natexlab{a}}).

\bibitem[{\citenamefont{Pierce and Manousakis}(1998)}]{Pierce98}
\bibinfo{author}{\bibfnamefont{M.}~\bibnamefont{Pierce}} \bibnamefont{and}
  \bibinfo{author}{\bibfnamefont{E.}~\bibnamefont{Manousakis}},
  \bibinfo{journal}{Phys. Rev. Lett.} \textbf{\bibinfo{volume}{81}},
  \bibinfo{pages}{156} (\bibinfo{year}{1998}).

\bibitem[{\citenamefont{Pierce and Manousakis}(1999)}]{Pierce99}
\bibinfo{author}{\bibfnamefont{M.}~\bibnamefont{Pierce}} \bibnamefont{and}
  \bibinfo{author}{\bibfnamefont{E.}~\bibnamefont{Manousakis}},
  \bibinfo{journal}{Phys. Rev. B} \textbf{\bibinfo{volume}{59}},
  \bibinfo{pages}{3802} (\bibinfo{year}{1999}).

\bibitem[{\citenamefont{Kwon and Whaley}(1999)}]{Kwon99}
\bibinfo{author}{\bibfnamefont{Y.}~\bibnamefont{Kwon}} \bibnamefont{and}
  \bibinfo{author}{\bibfnamefont{K.~B.} \bibnamefont{Whaley}},
  \bibinfo{journal}{Phys. Rev. Lett.} \textbf{\bibinfo{volume}{83}},
  \bibinfo{pages}{4108(4)} (\bibinfo{year}{1999}).

\bibitem[{\citenamefont{Kleinert}(2004)}]{Kleinert}
\bibinfo{author}{\bibfnamefont{H.}~\bibnamefont{Kleinert}},
  \emph{\bibinfo{title}{Path Integrals in Quantum Mechanics, Statistics,
  Polymer Physics, and Financial Markets}} (\bibinfo{publisher}{World
  Scientific Publishing Co. Pte. Ltd}, \bibinfo{address}{Singapore},
  \bibinfo{year}{2004}), \bibinfo{note}{3rd Edition}.

\bibitem[{\citenamefont{Pollock and Ceperley}(1987)}]{Pol87}
\bibinfo{author}{\bibfnamefont{E.~L.} \bibnamefont{Pollock}} \bibnamefont{and}
  \bibinfo{author}{\bibfnamefont{D.~M.} \bibnamefont{Ceperley}},
  \bibinfo{journal}{Phys.~Rev.~B} \textbf{\bibinfo{volume}{36}},
  \bibinfo{pages}{8343} (\bibinfo{year}{1987}).

\bibitem[{\citenamefont{Feynman}(1998)}]{Fey72}
\bibinfo{author}{\bibfnamefont{R.~P.} \bibnamefont{Feynman}},
  \emph{\bibinfo{title}{Statistical Mechanics}} (\bibinfo{publisher}{Perseus
  Books}, \bibinfo{year}{1998}).

\bibitem[{\citenamefont{Ceperley}(1995{\natexlab{b}})}]{Cep95}
\bibinfo{author}{\bibfnamefont{D.~M.} \bibnamefont{Ceperley}},
  \bibinfo{journal}{Rev.~Mod.~Phys.} \textbf{\bibinfo{volume}{67}},
  \bibinfo{pages}{279} (\bibinfo{year}{1995}{\natexlab{b}}).

\bibitem[{\citenamefont{Metropolis et~al.}(1953)\citenamefont{Metropolis,
  Rosenbluth, Rosenbluth, Teller, and Teller}}]{Metro53}
\bibinfo{author}{\bibfnamefont{N.}~\bibnamefont{Metropolis}},
  \bibinfo{author}{\bibfnamefont{A.~W.} \bibnamefont{Rosenbluth}},
  \bibinfo{author}{\bibfnamefont{M.~N.} \bibnamefont{Rosenbluth}},
  \bibinfo{author}{\bibfnamefont{A.~H.} \bibnamefont{Teller}},
  \bibnamefont{and} \bibinfo{author}{\bibfnamefont{E.}~\bibnamefont{Teller}},
  \bibinfo{journal}{J.~Chem.~Phys.} \textbf{\bibinfo{volume}{21}},
  \bibinfo{pages}{1087} (\bibinfo{year}{1953}).

\bibitem[{\citenamefont{Chakravarty et~al.}(1998)\citenamefont{Chakravarty,
  Gordillo, and Ceperley}}]{Chakravarty98}
\bibinfo{author}{\bibfnamefont{C.}~\bibnamefont{Chakravarty}},
  \bibinfo{author}{\bibfnamefont{M.~C.} \bibnamefont{Gordillo}},
  \bibnamefont{and} \bibinfo{author}{\bibfnamefont{D.~M.}
  \bibnamefont{Ceperley}}, \bibinfo{journal}{J. Chem. Phys.}
  \textbf{\bibinfo{volume}{109}}, \bibinfo{pages}{2123} (\bibinfo{year}{1998}).

\bibitem[{\citenamefont{Herman et~al.}(1982)\citenamefont{Herman, Bruskin, and
  Berne}}]{Herman82}
\bibinfo{author}{\bibfnamefont{M.~F.} \bibnamefont{Herman}},
  \bibinfo{author}{\bibfnamefont{E.~J.} \bibnamefont{Bruskin}},
  \bibnamefont{and} \bibinfo{author}{\bibfnamefont{B.~J.} \bibnamefont{Berne}},
  \bibinfo{journal}{J.~Chem.~Phys.} \textbf{\bibinfo{volume}{76}},
  \bibinfo{pages}{5150} (\bibinfo{year}{1982}).

\bibitem[{\citenamefont{Thijssen}(2000)}]{Thijssen}
\bibinfo{author}{\bibfnamefont{J.~M.} \bibnamefont{Thijssen}},
  \emph{\bibinfo{title}{Computational Physics}}
  (\bibinfo{publisher}{Cambridge}, \bibinfo{year}{2000}).

\bibitem[{\citenamefont{Ivanov et~al.}(2003)\citenamefont{Ivanov, Lyubartsev,
  and Laaksonen}}]{Ivanov03}
\bibinfo{author}{\bibfnamefont{S.~D.} \bibnamefont{Ivanov}},
  \bibinfo{author}{\bibfnamefont{A.~P.} \bibnamefont{Lyubartsev}},
  \bibnamefont{and}
  \bibinfo{author}{\bibfnamefont{A.}~\bibnamefont{Laaksonen}},
  \bibinfo{journal}{Phys. Rev. E} \textbf{\bibinfo{volume}{67}},
  \bibinfo{pages}{066710} (\bibinfo{year}{2003}).

\bibitem[{\citenamefont{Corso et~al.}(1996)\citenamefont{Corso, Pasquarello,
  and Baldereschi}}]{PP}
\bibinfo{author}{\bibfnamefont{A.~D.} \bibnamefont{Corso}},
  \bibinfo{author}{\bibfnamefont{A.}~\bibnamefont{Pasquarello}},
  \bibnamefont{and}
  \bibinfo{author}{\bibfnamefont{A.}~\bibnamefont{Baldereschi}},
  \bibinfo{journal}{Phys.~Rev.~B} \textbf{\bibinfo{volume}{53}},
  \bibinfo{pages}{1180} (\bibinfo{year}{1996}).

\bibitem[{\citenamefont{Kyl\"anp\"a\"a}(2006)}]{Ile06}
\bibinfo{author}{\bibfnamefont{I.}~\bibnamefont{Kyl\"anp\"a\"a}}, Master's
  thesis, \bibinfo{school}{Tampere University of Technology}
  (\bibinfo{year}{2006}).

\bibitem[{\citenamefont{Alexander and Coldwell}(2005)}]{Alexander05}
\bibinfo{author}{\bibfnamefont{S.~A.} \bibnamefont{Alexander}}
  \bibnamefont{and} \bibinfo{author}{\bibfnamefont{R.~L.}
  \bibnamefont{Coldwell}}, \bibinfo{journal}{Chem.~Phys.~Lett.}
  \textbf{\bibinfo{volume}{413}}, \bibinfo{pages}{253} (\bibinfo{year}{2005}).

\bibitem[{\citenamefont{Morse}(1929)}]{Morse29}
\bibinfo{author}{\bibfnamefont{P.~M.} \bibnamefont{Morse}},
  \bibinfo{journal}{Phys. Rev.} \textbf{\bibinfo{volume}{34}},
  \bibinfo{pages}{57} (\bibinfo{year}{1929}).

\bibitem[{\citenamefont{Lounila and Rantala}(1991)}]{Lounila91}
\bibinfo{author}{\bibfnamefont{J.}~\bibnamefont{Lounila}} \bibnamefont{and}
  \bibinfo{author}{\bibfnamefont{T.~T.} \bibnamefont{Rantala}},
  \bibinfo{journal}{Phys. Rev. A} \textbf{\bibinfo{volume}{44}},
  \bibinfo{pages}{6641} (\bibinfo{year}{1991}).

\bibitem[{\citenamefont{ter Haar}(1946)}]{terHaar}
\bibinfo{author}{\bibfnamefont{D.}~\bibnamefont{ter Haar}},
  \bibinfo{journal}{Phys.~Rev.} \textbf{\bibinfo{volume}{70}},
  \bibinfo{pages}{222} (\bibinfo{year}{1946}).

\bibitem[{\citenamefont{Kobus et~al.}(2005)\citenamefont{Kobus, Laaksonen, and
  Sundholm}}]{Pyykko}
\bibinfo{author}{\bibfnamefont{J.}~\bibnamefont{Kobus}},
  \bibinfo{author}{\bibfnamefont{L.}~\bibnamefont{Laaksonen}},
  \bibnamefont{and} \bibinfo{author}{\bibfnamefont{D.}~\bibnamefont{Sundholm}},
  \emph{\bibinfo{title}{A numerical hartree-fock program for diatomic
  molecules}} (\bibinfo{year}{2005}),
  \urlprefix\url{http://scarecrow.1g.fi/num2d.html}.

\end{thebibliography}

\end{document}